\documentclass[MSNbibl,nameyear,dvips]{arxstspdf}
\usepackage{flushend}
\usepackage{stfloats}
\usepackage{graphicx,url,breakurl}
\volume{29}
\issue{1}
\pubyear{2014}
\firstpage{26}
\lastpage{35}
\doi{10.1214/13-STS458} % kopijuoti is 'New paper accepted'
\begin{document}
\begin{frontmatter}

\title{How Bayesian Analysis Cracked the~Red-State, Blue-State Problem}%
% kai straipsnis turi susijusiu diskusiju ir rejoinder'iu
%rejoinder at \relateddoi{r}{10.1214/00-STSXXXX}.}
\runtitle{How Bayesian Analysis Cracked the Red-State, Blue-State Problem}

\begin{aug}
\author[a]{\fnms{Andrew} \snm{Gelman}\corref{}\ead[label=e1]{gelman@stat.columbia.edu}}
\runauthor{A. Gelman}

\affiliation{Columbia University}

\address[a]{Andrew Gelman is Professor,
Departments of Statistics and Political Science, Columbia University, New York, New York 10027, USA \printead{e1}.}

\end{aug}

% ABSTRACT
%
\begin{abstract}
In the United States as in other countries, political and economic
divisions cut along geographic and demographic lines. Richer \emph
{people} are more likely to vote for Republican candidates while poorer
voters lean Democratic; this is consistent with the positions of the
two parties on economic issues. At the same time, richer \emph{states}
on the coasts are bastions of the Democrats, while most of the
generally lower-income areas in the middle of the country strongly
support Republicans. During a research project lasting several years,
we reconciled these patterns by fitting a series of multilevel models
to perform inference on geographic and demographic subsets of the
population. We were using national survey data with relatively small
samples in some states, ethnic groups and income categories; this
motivated the use of Bayesian inference to partially pool between
fitted models and local data. Previous, non-Bayesian analyses of income
and voting had failed to connect individual and state-level patterns.
Now that our analysis has been done, we believe it could be replicated
using non-Bayesian methods, but Bayesian inference helped us crack the
problem by directly handling the uncertainty that is inherent in
working with sparse data.
\end{abstract}

% KEYWORDS

\end{frontmatter}

%s1 #&#
\section{Introduction}

Income and economic redistribution are central to electoral politics.
In the United States as in other countries, political and economic
divisions cut along geographic and demographic lines. Richer \emph
{people} are more likely to vote for Republican candidates while poorer
voters lean Democratic; this is consistent with the positions of the
two parties on economic issues. At the same time, richer \emph{states}
on the coasts are bastions of the Democrats, while most of the
generally lower-income areas in the middle of the country strongly
support Republicans. This geographic pattern is consistent with the
sense of a culture war between richer, more socially liberal
cosmopolitans and middle-class proponents of traditional American values.

These statistical patterns of voting at the individual and group level
are central to political debates about economic and social
polarization. During a research project lasting several years, we
resolved the statistical questions by fitting a series of multilevel
models to study the differences in voting between rich and poor voters,
and rich and poor states. We were using national survey data with
relatively small samples in some states, ethnic groups and income
categories; this motivated the use of Bayesian inference to partially
pool between fitted models and local data.

Previous, non-Bayesian analyses of income and \mbox{voting} had failed to
connect individual and state-level patterns. Typical analyses would be
either at the individual or the aggregate levels but not both. In the
studies that did model voting based on individual and geographic
characteristics, the focus was on estimating some particular regression
coefficient (or, more generally, on identification of some average
causal effect). Classical statistics tends to focus on estimation or
testing for a single parameter or low-dimensional vector, whereas
Bayesian methods work particularly well when the goal is inference
about a large number of uncertain quantities (in this case,
coefficients within each of the fifty states).\looseness=1

Now that our analysis has been done, we believe it could be replicated
using non-Bayesian methods: that is, with knowledge of the patterns we
have found, one could fit a simpler, non-Bayesian model to estimate the
interaction between individual and state incomes. However, one can also
view our fitting of a series of models as a form of exploratory data
analysis. It is only through active engagement with the data that we
got a sense of what to look for. Thus, the flexible generality of the
Bayesian approach facilitated our substantive research breakthrough here.

This is the opposite of the paradigm common in classical theoretical
statistics, of laser-like focus on identification of a single effect
and a concern with frequency properties of a prechosen statistical procedure.

Our Bayesian procedures are consistent with our political
knowledge---that is, having obtained our estimates, we and others have
been able to incorporate them into our understanding of income and
voting (as discussed in detail by Gelman et al., \citeyear{autokey8}, where we
consider various other factors including issue attitudes and
religiosity as individual-level predictors). But this sort of
theoretical coherence is not enough: social scientists are notoriously
adept at coming up with explanations to fit any set of supposed facts.
In addition, our model, as we have developed and extended it, fits the
data via graphical checks (for an example, see Figure~\ref{diagnostics},
which appears near the end of this article after we have
described the model), and, perhaps convincingly, has performed well in
external validation (in that we developed our models to fit to data
from the 2000 election and then they successfully worked for 2004, 2008
and 2012).

The real world impact of this work is twofold. First, we have
established that income is more strongly predictive of Republican
voting in poor states than in rich states, and that this difference has
arisen in the past two decades. Second, political scientists and
journalists now have a clearer view of the relation between social,
economic and political polarization. The political differences between
``red America'' and ``blue America'' are concentrated among the upper
half of the income distribution. By allowing us to model a pattern of
income and voting that varies across states, Bayesian analysis allowed
us to get a grip on this important political trend.

%s2 #&#
\section{Background}\label{sec2}

For the past fifteen years or so, Americans have been divided
politically into ``red states'' (the conservative, Republican-leaning
areas in the south and middle of the country) and ``blue states'' (the
more urban areas in the northeast and west coast, whose residents
consistently vote for Democrats). Here is the red-state, blue-state
paradox: since the 1990s, the poorer states have voted for conservative
Republicans while rich states favor liberal Democrats. This has
surprised political observers, given that Republicans traditionally
represent the rich with the Democrats representing the poor. And,
indeed, Republican candidates do about 20 percentage points better
among rich voters than among poor voters, a gap that has persisted for decades.

The red-state blue-state pattern became widely apparent in the
aftermath of the disputed 2000 presidential election, when television
viewers became all too familiar with the iconic electoral map: blue
states on the coasts and upper midwest voting for Al Gore, red states
in the American heartland supporting George Bush, and Florida colored
blank awaiting the decision of the courts.

The result has confused political observers on both sides of the
political spectrum. On the right came a much-discussed magazine article
by David Brooks, comparing Montgomery County, Maryland, the liberal,
upper-middle-class suburb where he and his friends live, to rural,
conservative Franklin County, Pennsylvania, a short drive away but
distant in attitudes and values, with ``no Starbucks, no Pottery Barn,
no Borders or Barnes \& Noble,'' plenty of churches but not so many
Thai restaurants, ``a lot fewer sun-dried-tomato concoctions on
restaurant menus and a lot more meatloaf platters.'' On the left,
Thomas Frank's bestselling \emph{What's the Matter with Kansas} (2004)
was widely interpreted to answer the question of why low-income
Americans vote Republican: ``For more than thirty-five years, American
politics has followed a populist pattern \ldots the average American,
humble, long-suffering, working hard, and paying his taxes; and the
liberal elite, the know-it-alls of Manhattan and Malibu, sipping their
lattes as they lord it over the peasantry with their fancy college
degrees and their friends in the judiciary.''

Here is a summary from \citet{Gel}:

\begin{quote}
Republicans, who traditionally represented America's elites, had
dominated in lower-income areas in the South and Midwest and in
unassuming suburbs, rather than in America's glittering centers of power.
What could explain this turnaround? The most direct story---hinted at
by Brooks in his articles and books on America's new, cosmopolitan,
liberal upper class---is that the parties simply switched, with the
new-look Democrats representing hedge-fund billionaires, college
professors and other urban liberals, and Republicans getting the votes
of middle-class middle Americans. This story of partisan reversal has
received some attention from pundits. For example, TV talk show host
Tucker Carlson said, ``Okay, but here's the fact that nobody ever, ever
mentions---Democrats win rich people. Over \$100,000 in income, you are
likely more than not to vote for Democrats. People never point that
out. Rich people vote liberal.'' And Michael Barone, the editor of the
Almanac of American Politics, wrote that the Democratic Party ``does
not run very well among the common people.'' But Tucker Carlson and
Michael Barone were both wrong \ldots obviously wrong, from the
standpoint of any political scientist who knows opinion polls.
Republican candidates consistently do best among upper-income voters
and worst at the low end. In the country as a whole and separately
among Whites, Blacks, Hispanics and others, richer Americans are more
likely to vote Republican. \ldots

Misconceptions about income and voting are all over the place in the
serious popular press. For example, James Ledbetter in \emph{Slate}
claimed that ``America's rich now tilt politically left in their
opinions.'' In the \emph{London Review of Books}, political theorist
David Runciman wrote, ``It is striking that the people who most dislike
the whole idea of healthcare reform---the ones who think it is
socialist, godless, a step on the road to a police state---are often
the ones it seems designed to help. \ldots Right-wing politics has
become a vehicle for channeling this popular anger against intellectual
snobs. The result is that many of America's poorest citizens have a
deep emotional attachment to a party that serves the interests of its
richest.'' No, no and no. An analysis of opinion polls finds, perhaps
unsurprisingly, but in contradiction to the above claims, that older
and high-income voters are the groups that most strongly oppose health
care reform.
\end{quote}

It has been difficult for political journalists to accept that richer
voters prefer Republicans while richer states lean Democratic.
At first this may appear to be a simple example of the ecological
fallacy: the correlation of income with Republican voting is negative
at the aggregate level and positive at the individual level.
And, indeed, part of the problem is a simple and familiar difficulty of
statistical understanding, associated with people assuming that
aggregates (in this case, states) have the properties of individuals.
This misunderstanding is particularly relevant from
a political perspective because the United States has a federal system
of government, with some policies determined nationally and others at
the state level. Thus, individual preferences and state averages are
both important in considering politics and policy in this country.

%s3 #&#
\section{Statistical Model and Bayesian Inference: Overview}

It turns out that the statistical story is more complicated too. Red
states and blue states do not only differ in their political
complexions; in addition, the relation between income and voting varies
systematically by state. In richer, liberal states such as New York and
California, there is essentially no correlation between income and
voting---rich and poor vote the same way---while in conservative states
such as Texas, the rich are much more Republican than the poor.
Political divisions by social class look different in red and blue America.

This key statistical part of our analysis is the estimation of the
relation between income and voting (later including religious
attendance and ethnicity as additional explanatory factors) separately
in each state. This is difficult because even a large national survey
will not have a huge sample size in all fifty states---and recall that
we are not merely estimating an average in each state but we are
attempting to estimate a regression or even a nonlinear functional
relationship. Political scientists armed with conventional statistical
tools sometimes try to get around this sample size problem by pooling
data from multiple years---but this would not work here because we are
also interested in changes over time.

The Bayesian resolution was a multilevel model allowing different
patterns of income and voting in different states. The model was built
on a hierarchical logistic regression but included error terms at every
level so that the ultimate fit was not constrained to fall along any
parametric curve. Because of the complexity of our model, it\vadjust{\goodbreak} was
necessary to check its fit by comparing data to posterior simulations.
Classical approaches---even classical multilevel models---would not
fully express the uncertainty in the fit. In contrast, our Bayesian
approach not only allowed us to fit the data, it also provided a
structure for us to consider a series of different models to explore
the data.

In general, estimating state-level patterns from national polls
requires two tasks: \emph{survey weighting} or adjustment for known
differences between sample and population (for example, surveys tend to
overrepresent women, whites and older Americans, while
underrepresenting young male ethnic minorities), and \emph{small-area
estimation} or regularized estimates for subsets where raw-data
averages would be too noisy.

In order to estimate the pattern of income and voting within each
state, we used the strategy of \emph{multilevel regression and
poststratification} (MRP), a general approach to survey inference for
subsets of the population that has two steps:
\begin{enumerate}[2.]
\item[1.] Use a multilevel model to estimate the distribution of the
outcome of interest (in this case, vote preference, among those people
who plan to vote in the presidential election) given demographic and
geographic predictors which divide the population into categories. Here
we start with 250 cells (5 income categories within each of 50 states);
a later model considers four ethnicity categories as well, and, more
generally, the analysis could categorize people by sex, age, income,
marital status, religion, religious attendance and so on, easily
leading to more cells than survey respondents. It is the job of the
Bayesian model to come up with a reasonable inference for the joint
distribution of the Republican vote share within whatever categories
are included.
\item[2.] Poststratify to sum the inferences across cells. For example, the
estimated percentage of support for Obama among Hispanics in the
midwest is simply the weighted average of his estimated support within
each of the relevant poststratification cells (in the
ethnicity/income/state model, this would be one cell for each income
category within each midwestern state). The weights in this weighted
average are simply the number of voters in each cell, which we can get
from the U.S. Census. (To obtain voter weights is itself a two-stage
process in which we first take the number of adult Americans in each
cell, then multiply within each cell by the proportion of adults who
voted, as estimated from a multilevel logistic regression fit to a
Census post-election survey that asks about voting behavior; again, see
\cite*{GhiGel13}, for details.)
\end{enumerate}
%
%(In sampling jargon, \emph{strata} are defined based on the design of
%the survey---a stratified design has separate sampling within each
%stratum---whereas \emph{post-strata} are chosen based on the analysis.
%This is why our method is MRP and not MRS.)

It is clear how MRP fits in with Bayesian statistics: the number of
observations per cell is small, so our problem is one of small-area
estimation (\cite{FayHer79}), hence, it makes sense to partially
pool inferences, averaging local data and a larger fitted regression
model. Bayesian inference is a well-recognized tool for combining local
information with predictions from a stochastic model (\cite{ClaKal87}).

But it may be less obvious how our method connects with the vast
literature on survey weighting, a field that traditionally draws a
strong distinction between ``model-based'' procedures such as Bayesian
or even likelihood methods that posit a probability model for the data
and ``design-based'' inference which leave data unmodeled and apply a
probability distribution only to the sampling process. The connection
was made clear by Little (\citeyear{Lit91}, \citeyear{Lit93}), who showed how model-based
inference fits in a larger design-based framework (or, conversely, how
design-based inferences are possible within a larger probability
model). Little's key insight is centered on the \emph
{poststratification identity}:
\[
\theta=\frac{\sum_j N_j \theta_j}{\sum_j N_j}, %
\]
where $\theta$ is some aggregate quantity of interest (for example, the
estimated support for Obama among Hispanics in the midwest), $j$'s are
the cells within this aggregate, $N_j$ is the population size of each
cell (in our case, obtained from the Census), and $\theta_j$ is the
(unknown) population quantity with the cell.

As noted, the above equation is a tautology. Its connection to
statistical inference comes in the inferences for the $\theta_j$'s.
Assuming simple random sampling within cells (the implied basis for
classical survey weighting), one can estimate the $\theta_j$'s through
simple raw cell means (statistically inefficient if sample sizes are
small) or more effectively via regression modeling which quickly leads
to Bayes if the number of cells is large and the model is realistically
complex. The information that would go into classical survey weights
instead enters our MRP calculations through the population sizes $N_j$.
This is important: you can't get something for nothing, and the cost of
our poststratified lunch is the array of population numbers $N_j$.

MRP combines long-existing ideas in sample surveys but has become
recently popular as a way to learn about state-level opinions from
national polls (\cite*{GelLit97}; \cite*{LaxPhi09N1,LaxPhi09N2}),
perhaps as a result of increasing ease of handling large data sets as
well as improvements in off-the-shelf hierarchical modeling tools. In
many political science applications, state averages are of primary
interest, and we estimate opinion in within-state slices (for example,
white women aged 30--44 in Missouri) only because we feel we need to,
in order to adjust for differential nonresponse. We fit the multilevel
model to get reasonable inferences within all these cells but then
immediately poststratify to get state-level estimates. All these steps
are needed---a simple Bayesian analysis of state-level data would fail
to adjust for known demographic differences between sample and
population. Modern surveys have large problems with
nonrepresentativeness and some sort of adjustment is necessary to match
the population. MRP forms a bridge between Bayesian inference (so
flexible and powerful for estimating large numbers of parameters and
making large numbers of uncertain predictions at once) and classical
survey adjustment (given that real surveys can be clearly
nonrepresentative of the population). This latter step is crucial in
many applications in which data are combined from many disparate surveys.

In the Red State Blue State project, MRP plays a slightly different
role. Here we actually are interested in categories within a state
(initially, the five income categories; later, voters cross-classified
by income, education and religious attendance). The poststratification
is less important here (although it does come up: after we sum our
inferences over cells within each state, we adjust our predictions of
state-level averages to line up with actual recorded vote totals, a
completely reasonable step given this additional information separate
from the survey data). What is relevant for the present discussion is
that our method harnesses the power of Bayes within a framework that
accounts for concerns specific to survey sampling.

%s4 #&#
\section{Statistical Model and Bayesian Inference: Details}

We fit our models separately to pre-election poll data from 2000, 2004
and 2008, with about 20--40,000 respondents in each year. This sample
size is large enough for us to estimate variation among states but not
so large that we could just estimate each state's pattern using its own
data alone. An intermediate approach would be to combine similar
states, although one would not want to combine completely, and it would
then make sense to fit a regression model to determine which states to
combine, and also set some rule based on sample size to decide how much
to pool each state \ldots and this all leads to a multilevel regression.

For the purposes of learning about opinion from a sample, the
multilevel model is a way to obtain estimates for mutually exclusive
slices of the population (and implicitly corresponds to the assumption
that the respondents being analyzed are a simple random sample within
each cell). From the perspective of statistical inference, however, our
model is simply a hierarchical regression with discrete predictors.
Thus, if we want to perform inference for 4 ethnicities $\times$ 5
income categories $\times$ 50 states, we just need to include
predictors for ethnicities, income levels and states (along with
various interactions), and perform inferences for the vector of
regression coefficients, and inferences for the 1000 cells just pop out
as predictions from the fitted regression model.

The most basic form of the model is a varying-intercept logistic
regression of survey responses:
\[
\operatorname{Pr}(y_i=1)=\operatorname{logit}^{-1}(
\alpha_{j[i]} + X_i\beta), %
\]
where:
\begin{itemize}
\item $y_i=1$ if respondent $i$ intends to vote for the Republican
candidate for president or 0 if he or she supports the Democrat (with
those expressing no opinion excluded from the analysis),
\item$\alpha_{j[i]}$ is a varying intercept for the state $j[i]$ where
the respondent lives (that is, $j[i]$ is an index taking on a value
between 1 and 50),
\item$X_i$ is a vector of demographic predictors (indicators for
state, age, ethnicity, education and some of their interactions, and
also income, discretized on a scale of $-2,-1,0,1,2$), and $\beta$ is a
vector of estimated coefficients.
\end{itemize}
The intercepts $\alpha_j$ are themselves modeled by a regression:
\[
\alpha_j \sim\mathrm{N}\bigl(W_j\gamma,
\sigma^2_{\alpha}\bigr), %
\]
where:
\begin{itemize}
\item$W_j$ is a vector of state-level predictors (including average
income of the residents of the state, Republican vote share in the
previous presidential election and indicators for region of the country),
\item$\gamma$ is a vector of state-level coefficients, and
\item$\sigma_{\alpha}$ is the standard deviation of the unexplained
state-level variance.
\end{itemize}
We completed the Bayesian model by assigning to the otherwise unmodeled
parameters $\beta$, $\gamma$, $\sigma_{\alpha}$ a uniform prior
distributions: in retrospect, not the best choice (we do in fact have
prior information on these quantities, starting with results from the
model fit to earlier elections) but enough to give us reasonable
results. As this work goes forward we plan to think harder about
hyperprior distributions and additional levels of the hierarchy such as
building in time-series models.\vadjust{\goodbreak}

%f1 #&#
\begin{figure*}

\includegraphics{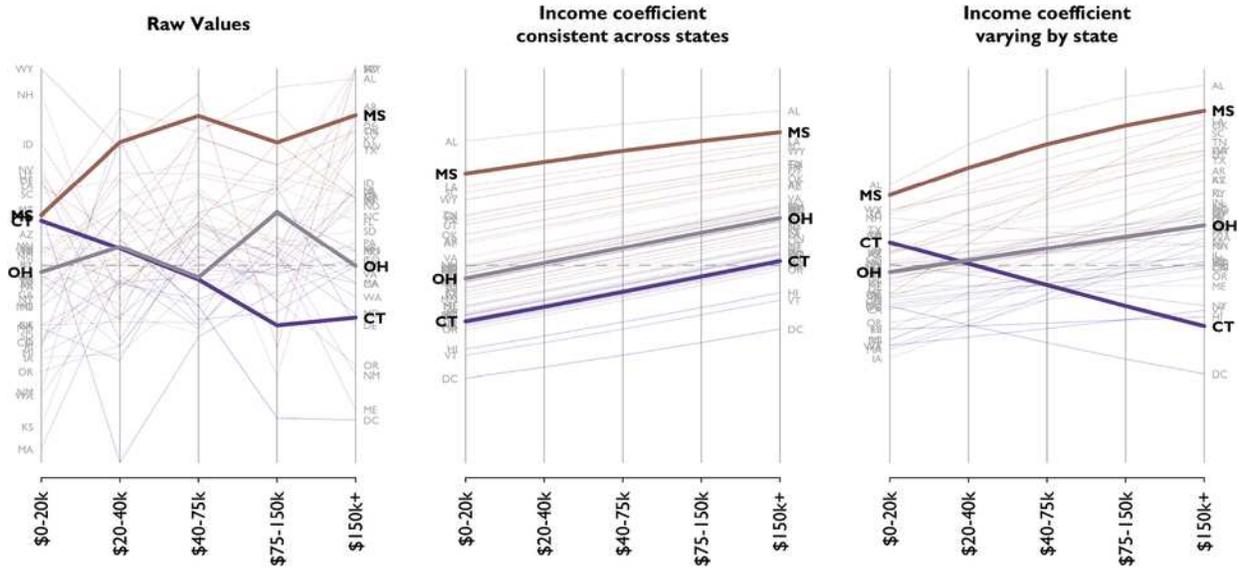}

\caption{The evolution of a simple model of vote choice in the
2008 election for state $\times$ income subgroups, non-Hispanic
whites only.  The colors come from the 2008 election, with darker
shades of red and blue for states that had larger margins in favor
of McCain or Obama, respectively. The first panel shows the raw data;
the middle panel is a hierarchical model where state coefficients
vary but the (linear) income coefficient is held constant across states;
the right panel allows the income coefficient to vary by state. Adding
complexity to the model reveals weaknesses in inferences drawn from
simpler versions of the model. The poorest state (Mississippi), a
middle-income state (Ohio), and the richest state (Connecticut)
are highlighted to show important trends.
From Ghitza and Gelman (\citeyear{GhiGel13}).}\label{logisticmodels}
\end{figure*}

The varying-intercept model above fails because it assumes a constant
relation between income and voting across states. Actually, the data
show that income is much more highly correlated with Republican voting
in some states than others. We fit this pattern using a model in which
the intercept and the coefficient for individual income varies by
state. The two varying coefficients within each state are then
themselves modeled given state-level predictors and with a $2\times2$
covariance matrix for the state-level errors. (We coded income as $-2$
to 2 rather than 1--5 so that the joint distribution of intercept and
slope would be easier to model, following standard practice in
regressions with interactions.) Figure~\ref{logisticmodels} shows the
models with constant slope and then with income coefficients varying by
state. This new model fit reasonably well but we further elaborated it
by adding varying coefficients for each income category, thus allowing
a nonlinear relation (on the logistic scale) of income and vote
preference that could vary by state and ethnicity. Income is included
in this regression in three ways at once, but because of the
hierarchical Bayesian model there is no multicollinearity problem.

Other versions of the model include additional individual-level
predictors such as age, education and religious attendance. For some
polls that are ``self-weighting'' or approximately so---this refers to
surveys where adjustments are made within the sampling process to
minimize demographic differences between sample and population---we
also sometimes fit models with \emph{fewer} individual predictors.
Ideally it makes sense to include important predictors such as sex, age
and ethnicity in the poststratification to correct for sampling bias
and variance in these dimensions, but for simplicity in computation and
analysis we have also fit models including only income as a
respondent-level variable.\looseness=1

\begin{figure*}

\includegraphics{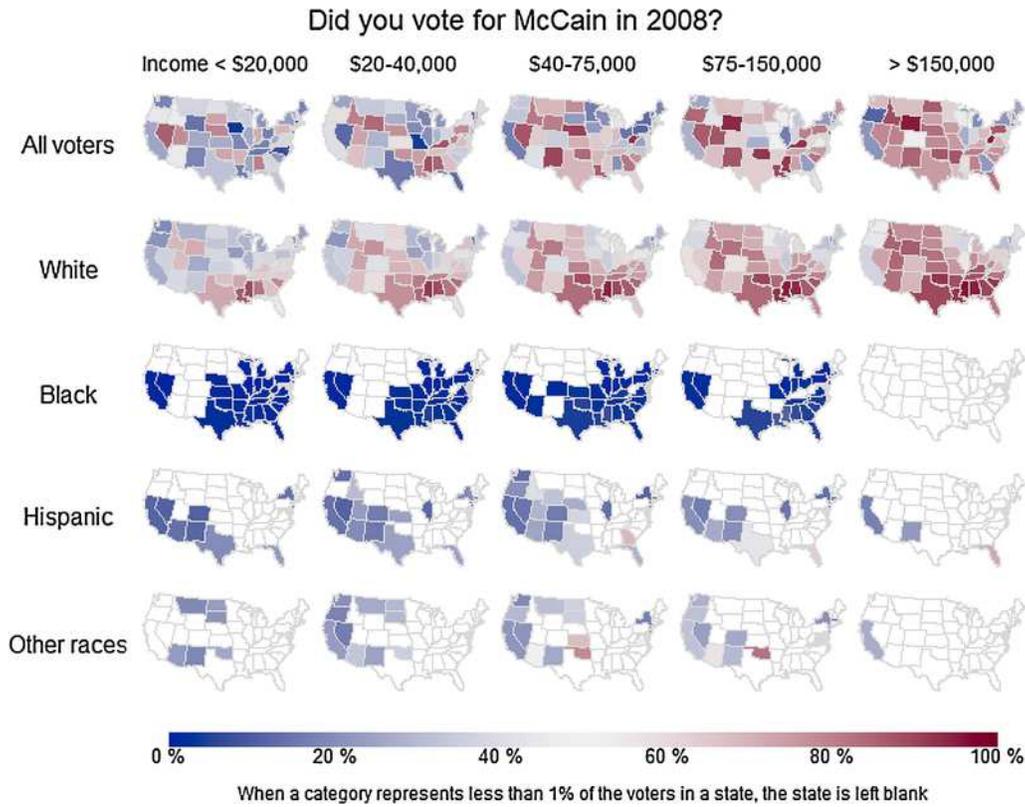}

\caption{Estimated two-party vote share for John McCain in the 2009 presidential election, as
estimated using multilevel modeling and poststratification
from pre-election polls.  Figure \protect\ref{diagnostics}
displays some graphical diagnostics for comparing this fitted
model to data.  From Gelman, Lee and Ghitza (\citeyear{GelLeeGhi10N1}).}\label{fullmodel}
\end{figure*}

The different pieces of the Bayesian predictive model for vote
preferences connect in different ways to our statistical and
substantive goals. Adjustments for sex, age, ethnicity and education
correspond to survey weighting for these variables to correct for
important known differences between sample and population. Including
individual income as a predictor serves the goal of comparing the votes
of rich and poor within states, while including state income as a
group-level predictor allows us to compare rich and poor states.
Finally, the varying-intercept model for state with its error term
allows unexplained variation among states, which is crucial because we
know that states vary in many other ways beyond that predicted by state
income levels. The final model can be written in the form
\[
\mathrm{E}(y) \mbox{ within cell }j = \operatorname{logit}^{-1}(W_j
\gamma), %
\]
where $j$ indexes the poststratification cell (states $\times$
demographic variables), $W$ is a\vadjust{\goodbreak} matrix of indicator variables, and
$\gamma$ is a vector of logistic regression coefficients which
themselves are modeled hierarchically, with batches of main effects and
interactions. In the Bayesian analysis, posterior simulations are
obtained on $\gamma$, which in turn induces a posterior distribution on
the cell means, which are then combined by weighting with census
numbers to obtain estimates for any subsets of the population.
Figure~\ref{fullmodel} shows the resulting estimates of vote
preferences by state, ethnicity and income for the 2008 presidential election.

%f2 #&#

The poststratification step points to a difference between our Bayesian
solution and traditional \mbox{statistical} analyses. Even our basic model had
many parameters but none of them mapped directly to our summaries of
interest. To obtain the relation between income and voting within a
state, we did not look at the coefficient for the income predictor.
Rather, we used our model to estimate opinion in each
poststratification cell and then summed up to infer about each income
category within each state. Similarly, we compare rich and poor states
not by focusing on the coefficient of state income in the group-level
regression but by using MRP to estimate the slope in each of the 50
states and then plotting the estimates vs. state income. The individual
and state-level income coefficients are relevant to the model, but our
ultimate inferences are constructed from pieced-together predictions.
This sort of simulation-based inference may seem awkward to
classically-trained statisticians but its flexibility makes it ideal
for problems in political science where we are interested in studying
variation rather than in estimating some sort of universal constant
such as the speed of light. In addition, simulation-based estimates can
be directly and easily expressed on the probability scale; there is no
need to try to interpret log-odds or logistic regression
coefficients.\looseness=1

%s5 #&#
\section{Gains from Bayes}

Income and voting had been studied by political scientists for decades,
but it was only through Bayesian methods that we were able to discover
the different patterns of income and voting in rich and poor states, an
important and exciting pattern that had never been noticed before. At a
technical level, our approach also accounted for the design of the
survey data by adjusting for demographic factors that were used in
survey weighting.

%f3 #&#
\begin{figure*}[t!]

\includegraphics{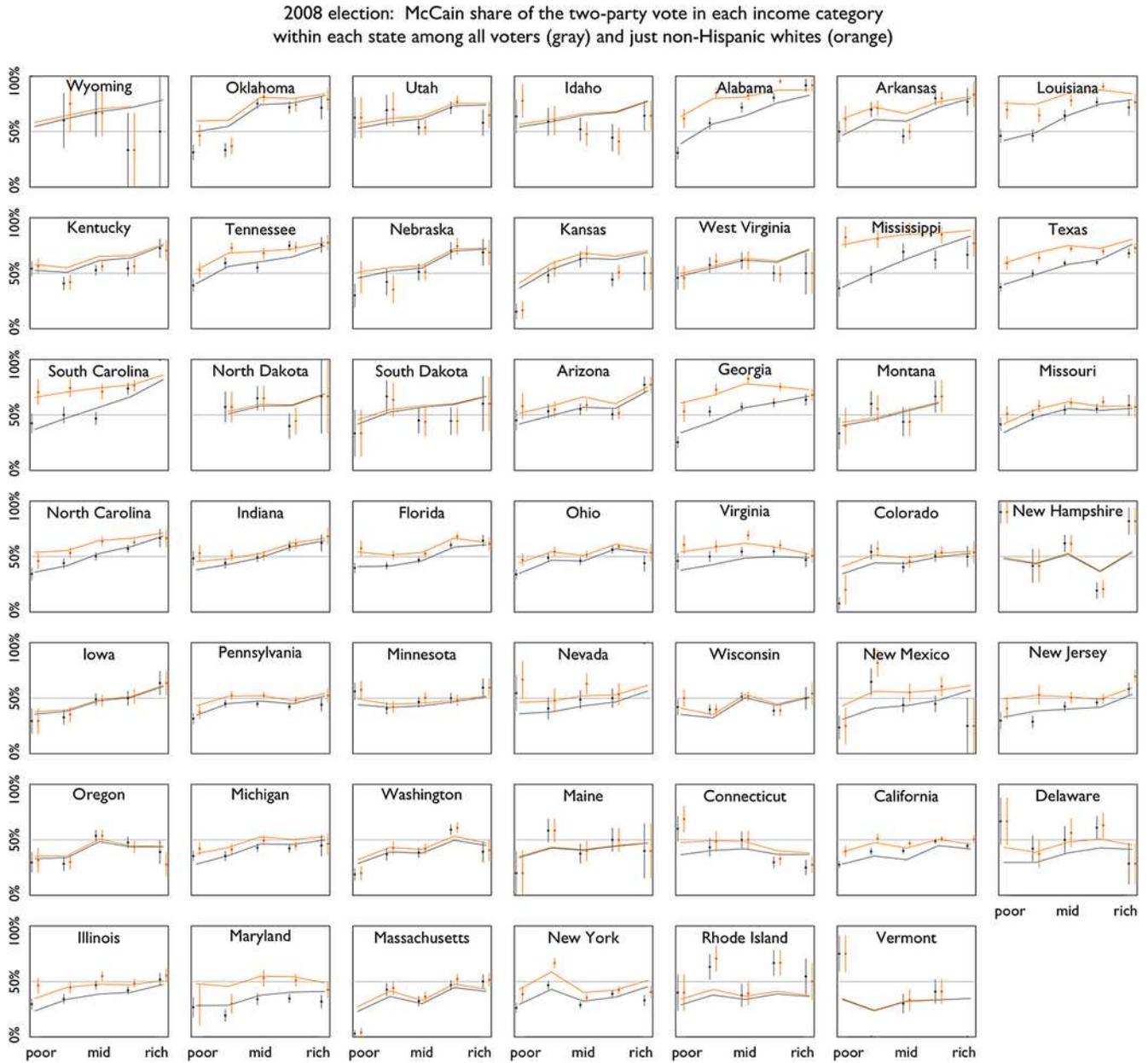}

\caption{Share of the two-party vote received by John McCain in
each income category within each state among all voters (gray)
and non-Hispanic whites (orange).  Dots are weighted averages
from pooled survey data from the five months before the election;
error bars show $\pm 1$ standard error bounds.  Curves are estimated
using multilevel models and have a standard error of about 3\% at each
point.  States are ordered in decreasing order of McCain vote share
(Alaska, Hawaii and the District of Columbia excluded).  From Ghitza
and Gelman (\citeyear{GhiGel13}).}\label{diagnostics}
\end{figure*}

Often the key to a statistical method is not what it does with the data
but, rather, what data it allows one to use. MRP combines design-based
and\vadjust{\goodbreak} model-based inference and can handle data from multiple surveys as
well as census totals on demographics. As always, Bayesian inference
works well with models with large numbers of parameters, allowing
adjustment for many factors, which is another way of including more
information in the inferential procedure. The complexity of the
resulting inferences make it particularly important to graphically
check the fit of model to data, as demonstrated in
Figure~\ref{diagnostics}.\looseness=1

We consider any multilevel model here to be\break Bayesian, even if it takes
the form of a classical mixed-model in which the fixed effects and
hierarchical \mbox{variance} parameters are estimated using marginal maximum
likelihood. In the context of using surveys to estimate public opinion
in geographic and demographic slices of the population, the inferences
for quantities of interest are constructed from the estimated joint
predictive distribution of the cell expectations. This summary of
knowledge in the form of a probability distribution is the essence of
Bayesian inference.

That said, we believe that an analysis just as good as ours could be
constructed entirely using non-Bayesian methods. It would require a lot
of extra work (for us) but it should be possible. In fact, many of the
patterns we discovered (most notably, that income predicts Republican
voting better in rich states than in poor states, and that religious
attendance predicts Republican voting better among rich than poor
voters) appear directly in the raw data---if you know to look for them.
In that sense, multilevel Bayesian modeling (adapted to the sample
survey context using poststratification) can be considered as an
elaborate form of exploratory data analysis, giving us the chance to
see patterns of complex interactions that are in the data but would not
appear in simple regression models.

The key pieces in the Bayesian inference were: (a)~weighted averages
for small-area estimation;\break (b)~poststratification, which detached the
modeling stage of the analysis from the inferences for quantities of
interest; (c) state-level predictors, which gave us reasonable
estimates even for small states; (d)~individual-level income included
as a continuous and discrete variable at the same time, allowing a
nonparametric form for the income--voting relation but partially pooling
to linearity; (e) and flexibility in modeling, letting us see the data
and examine the problem from many different angles without the burden
of requiring a fully-specified model. In standard statistical
theory---Bayesian or otherwise---a model is either already built or is
one of some discrete class of candidate models. In this sort of applied
exploration, however, the model is always evolving, and it is helpful
to have a statistical and computational framework in which we can
explore different possibilities. The Bayesian framework is particularly
open-ended in that adding complexity to the model is just a matter of
adding parameters in the joint posterior distribution.

%s6 #&#
\section{Moving Forward}

Our book that built upon the analysis described above has changed how
journalists and political professionals think about the social and
political bases of support for America's two major political parties.
For example, in \citeyear{Fru09}, political journalist and former presidential
speechwriter David Frum described our book as ``must reading'':
\begin{quote}
At first glance, American voting seems
topsy-turvy. Super-wealthy communities\break like
Beverly Hills, Aspen, and the Upper East Side of
Manhattan vote Democratic. Meanwhile, Appalachia and Alaska are
becoming ever more Republican. Republicans accuse the Democrats of
``elitism.'' Liberals wonder ``what's the matter with Kansas'' and
suspect low-income voters are either gullible or racist. Gelman
deconstructs the paradox. \ldots Most of us have the notion that
issues such as abortion, same-sex marriage and immigration divide a
more liberal, more permissive elite from a more traditionalist voting
base: Bob Reiner vs. Joe the Plumber. Not so, says Gelman, and he has
numbers to prove it. Downmarket voters are bread-and-butter voters. It
is upper America that is divided on social issues: a more permissive,
more liberal elite in the Northeast and California and a more
religious, more conservative elite in the South and Midwest. It's not
Hollywood vs. Wassila. It's Hollywood vs. the wealthy suburbs of Dallas
and Houston and Atlanta.
\end{quote}
This new understanding is, we hope, replacing the more simplistic
attitudes of rich Democrats and down-to-earth Republicans as expressed
by various pundits in Section~\ref{sec2} of this article.

The success of our red-state--blue-state analysis also motivated
ourselves and others to apply MRP in a variety of settings to
understand local attitudes and to integrate demographic and geographic
modeling in social science, on topics related to health care
(\cite{GelLeeGhi10N2}), capital punishment (\cite{ShiGel}),
gay rights (\cite{LaxPhi09N1}) and more general questions of the
relation between state-level opinion and state policies (\cite{LaxPhi12}). To the extent that public opinion can be estimated at
the state level and this is done for topical issues, this can inform
public debate, as, for example, with the recent Senate vote on the
Employment Nondiscrimination Act (\cite{LaxPhi13}).

On a more methodological level, many statistical challenges remain with
MRP, most notably how to build and compute models with many predictive
factors (age, ethnicity, education, family structure, \ldots) and
correspondingly huge numbers of interactions, how to visualize such
model fits, and how to poststratify on characteristics such as
religious attendance that are not known in the population. More
generally, our increasing ability to fit large statistical models puts
more of a burden on checking and understanding these models. Given that
a mere two-way model of income and state turned out to be complicated
enough to require a multi-year research project, we anticipate new
challenges in digesting larger models that allow more accurate
inferences from sample to population.

\section*{Acknowledgments}

Partially supported by the National Science Foundation and Institute of
Education Sciences.
We thank Yair Ghitza and two reviewers for helpful comments, and
Christian Robert and Kerrie Mengersen for organizing this special issue.

%suskaldyti doi

% imsref loaded by audrone.aklyte, 2014-02-14 09:48:34
% imsref loaded by audrone.aklyte, 2014-02-14 10:39:43
% imsref loaded by audrone.aklyte, 2014-02-14 10:41:46
% imsref loaded by audrone.aklyte, 2014-02-14 10:45:02
%
% imsref loaded by audrone.aklyte, 2014-04-01 08:57:44

\end{document}